\documentstyle[11pt,aaspp4]{article}

\newcommand{\beq}{\begin{equation}}
\newcommand{\eeq}{\end{equation}}
\newcommand{\lsim}{\lesssim}
\newcommand{\gsim}{\gtrsim}
\def\etal{{\it et al.}~}
\begin{document}

\title{The Cosmic Baryon Budget}  

\author{M. Fukugita}
\affil{Institute for Advanced Study, Princeton, NJ 08540, USA,
and\\ Institute for Cosmic Ray Research, University of Tokyo,
Tanashi, Tokyo 188, Japan}
\authoremail{fukugita@icrr.u-tokyo.ac.jp}

\author{C. J. Hogan} 
\affil{Departments of Astronomy and Physics, University of Washington,
Seattle, WA 98195, USA}
\authoremail{hogan@centaurus.astro.washington.edu}

\author{P. J. E. Peebles}
\affil{Joseph Henry Laboratories, Princeton University,
Princeton, NJ 08544, USA}
\authoremail{pjep@pupgg.princeton.edu}
 
\begin{abstract}
We present  an estimate of the global budget of
baryons in all states, with conservative estimates of the
uncertainties, based on all relevant information we have been
able to marshal. Most of the baryons today are still in the 
form of ionized gas, which contributes a mean density uncertain 
by a factor of about four.   
Stars and their remnants are a relatively minor component,
 comprising for our best-guess plasma density
 only about $17\%$ of the baryons, while
populations contributing most of the blue starlight comprise
less than 5\%. The formation of galaxies and of stars within
them  appears to be a globally inefficient process.
The sum over our budget, expressed as a fraction
of the critical Einstein-de~Sitter density, is in the range 
$0.007\lsim\Omega _B\lsim 0.041$,
with a best guess $\Omega _B\sim 0.021$ (at Hubble constant
70~km~s$^{-1}$~Mpc$^{-1}$). The central value agrees with the
prediction from the theory of light element production and
with measures of the density of intergalactic plasma at redshift
$z\sim 3$. This apparent concordance 
suggests we may be  
close to a complete survey of the major states of the baryons. 
\end{abstract}

\section{Introduction}

The evolution of structure in the expanding universe has
redistributed the baryons from a nearly smooth plasma at the time
of light element nucleosynthesis to a variety of states ----
condensed, atomic, molecular, and plasma --- in clouds of gas and 
dust, planets, stars, and stellar remnants, that are arranged in 
 the galaxies, in groups and clusters of galaxies, and in between.
The  amounts of baryons in each state
and form at low redshift can be compared  to what is
observed at higher redshift and to the total cosmic
abundance predicted by the theory of element production in  
the early universe.
Budget estimates must be informed by ideas on how
structure evolves as well as by the observations, and the budget 
in turn is a test of these ideas. Knowledge of the baryon budget
is an essential boundary condition for the analysis of how
structure formed and of the nature of cosmic dark matter.  

Advances in observations allow reasonably sound estimates of 
the amount of baryons present in a considerable variety of forms. 
In a previous discussion (Fukugita,
Hogan, \& Peebles 1996), we presented a picture for cosmic
evolution suggested by the results of budget
estimates.  In this paper we give details of the budget
calculations and update them using our choices for
the current best knowledge 
relevant to the calculation.
Persic \& Salucci (1992), Gnedin \& Ostriker (1992), and  Bristow \&
Phillipps (1994)  estimate baryon
abundances with different emphases; the results are compared in
\S 5.1. 

The main focus of this paper is the baryon budget at low
redshift; our accounting is presented in \S~\ref{section2}. The
low redshift budget can be compared to the situation at $z\sim 3$,
where quasar absorption lines allow a comprehensive accounting of
the diffuse components. We comment on this in \S~\ref{section3}. The
budgets are compared to the constraint 
 from light element production in \S~\ref{section4},
and the implications for galaxy and structure formation 
are summarized in \S~\ref{section5}.

We write Hubble's constant as
\beq H_o =100h\hbox{ km s}^{-1}\hbox{ Mpc}^{-1}.\label{eq:1c}\eeq
Where not explicitly written we use solar units for mass and
luminosity and megaparsecs for the length unit. 

\section{The Baryon Budget at $z\approx 0$}\label{section2}

\subsection{Stars and Remnants in Galaxies}\label{section2.1}

Stars in high surface density galaxies are the most prominent
place for baryons. 
For our purpose it is reasonable to imagine the galaxies contain 
two distinct stellar populations, an old spheroidal component
and a disk component consisting of generally younger stars,
with a mix depending on galaxy type.
Elliptical galaxies lack a significant disk component, and
irregular (Im) galaxies are in the  opposite extreme, having
small or insignificant bulges. The formation of these two
components seems to follow different histories, so it is
appropriate to count baryons in stars divided into these
two categories. The mass density for each component is obtained
as \beq\rho({\rm sph, disk})={\cal L}_B\cdot f_B({\rm sph, disk})\cdot
\langle M/L_B\rangle_{\rm sph, disk}, \label{eq:1a}\eeq
where ${\cal L}$ is the mean luminosity density, $f_B$ is the
fraction of the luminosity density produced by the spheroid or disk
component,
and $\langle M/L_B\rangle$ is the mass-to-light ratio for each component
including the stars and star remnants.
The suffix $B$ refers to luminosities measured in the $B$ band,
our choice for a standard wavelength band. 
Although the stellar population of irregular galaxies is similar to 
that of the disk component, we treat the former separately to
emphasize its distinct role in the luminosity density at 
low redshift and a slightly smaller mass-to-light ratio
because of its younger mean age.

\subsubsection{Luminosity Fractions}

We consider first the luminosity fractions $f$ in
equation~(\ref{eq:1a}). 
In Table~\ref{table1} we show the bulge fraction of galaxies in
luminosity, $\kappa=L_{\rm bulge}/(L_{\rm bulge}+L_{\rm disk})$,
for each  morphological type,
 from the bulge-disk decomposition of Kent (1985) based on his 
accurate Gunn 
$r$ band CCD photometry. We take the median of the distribution
presented by Kent. 
The bulge fraction in $r$ band is converted into that in the
$B$ band using the color transformation law $B-r=1.19$
for elliptical galaxies and bulges (Fukugita, Shimasaku
\& Ichikawa 1995). The values of $\kappa$ thus obtained in the
$B$ band are consistent with the result of Whitmore and Kirshner
(1981), although the latter sample is not as large as that of
Kent and the decomposition, based on photography, may be less
accurate. The bulge fraction is well correlated with
the morphology (Morgan 1962; Dressler 1980; 
Meisels \& Ostriker 1984; Kent 1985), 
but the morphologies at given bulge fraction overlap 
neighboring classes.

Row (3) in Table~\ref{table1} is our adopted distribution of
morphologies, expressed as the fraction $\mu _t$ of the
mean luminosity density contributed by galaxies of morphological
type $t$. Postman \& Geller (1985) 
give E:S0:S+Irr=0.12:0.23:0.65. This ratio is consistent
with the Tinsley mix (Tinsley 1980), which subdivides the spiral
fraction into further subclasses, while E and S0 are not
separated. The luminosity fractions in equation~(\ref{eq:1a}) are 
obtained by combining the two results:
\beq
	f_B({\rm sph}) = \sum\mu _t\kappa _t =0.385,\qquad
	f_B({\rm disk}) = \sum\mu _t(1-\kappa _t)=0.549,\qquad
	f_B({\rm irr}) = \mu _{\rm irr}=0.061.
		\label{eq:1}
\eeq
The sum in the second expression is from SOs through
spirals. Because the mass in S0 discs is so small
we make little error in treating the disks of spirals
and S0s as the same population. 
The separation into each morphological type is ambiguous, and the 
classification depends much on the author. To model such error,
we recalculate the fractions $f$ after reclassifying 1/3 of the
galaxies in each type systematically to the neighboring later
type, or else to the neighboring earlier type. This produces the
range (0.324--0.459):(0.599--0.499): (0.076--0.041). When we use
a morphological fraction or a bulge-to-disk 
ratio taken from other authors, we find the
luminosity fractions $f$ fall within this range in most cases.
In particular these estimates, when translated into the $V$ pass
band, agree with the survey of
Schechter and Dressler (1987).
This spread of values of the $\mu _t$ is taken into account in
our assignment of errors in estimates of the mean mass density.

\subsubsection{The Mean Luminosity Density}

Recent measurements of the mean luminosity density contributed by
high surface brightness galaxies at moderately low
redshift are summarized in Table~\ref{table2}. There is a
substantial spread of estimates of the  parameters
$\phi^*$ and $M^*$ in the Schechter function,
but the total luminosity density,
${\cal L} =\phi^*L^*\Gamma(\alpha+2)$, is better defined:
most values are in the range 
${\cal L}_B=(1.8-2.2)\times 10^8h L_\odot{\rm Mpc}^{-3}  $.
The exception is the significantly lower number from the APM-Stromlo
survey (Loveday et al. 1992). We adopt 
\beq
   {\cal L}_B=(2.0\pm 0.2)\times 10^8hL_\odot{\rm Mpc}^{-3},
 \qquad (M/L_B)_{\rm crit} = (1390\pm 140)h.\label{eq:1d}
\eeq
The second expression is the corresponding value of the
mass-to-light ratio in the Einstein-de~Sitter model.

\subsubsection{Mass-to-light Ratio of the Spheroid Component}

The mass-to-light ratio of   stars and 
their remnants in spheroid populations can be inferred from
star population synthesis calculations, assuming the stars are
close to coeval (so that the integrated $B$ luminosity is not much affected
by recent episodes of star formation)
and the spheroid stars in our neighborhood are representative.
Within these assumptions the most serious uncertainty
in the calculation is the lower cut off
of the initial mass function (the IMF). In the past the cutoff was 
somewhat arbitrary, between 0.1$M_\odot$ and 0.2$M_\odot$, and
the choice seriously affected  
the estimate of $M/L$. Thus for the Salpeter IMF 
$M$ differs by 24\% between  these two 
choices, while $L$ is virtually unaffected. This uncertainty
for the local IMF has been removed by the recent advance 
in observation of M subdwarf star counts with the Hubble
Space Telescope: it has been shown that the IMF of the
local Galactic stars 
has a turnover at about 0.4$M_\odot$ (Gould, Bahcall \& Flynn 1996;
hereafter GBF), and the integral over the GBF mass 
function converges without an arbitrary cut off. For $M>M_\odot$
the GBF mass function agrees well with the Salpeter function. 
Using the GBF mass function for $M<M_\odot$, instead of the 
Salpeter mass function with $x=1.35$ (where $x$ is the index of the
IMF $dn/dM\propto M^{-(1+x)}$), 
the mass integral is 0.70 times that obtained with the Salpeter mass
function cut off at 0.15$M_\odot$. 
The light in the blue-visual bands contributed by stars less
massive than one solar mass is 
small: from Table 4 of Tinsley \& Gunn (1976), we estimate that
the $V$ light integral decreases only 2\% with the use of the GBF 
mass function.  Therefore, the mass-to-light ratio with the GBF mass
function is $(M/L_V)_{\rm GBF}=0.72(M/L_V)_{\rm Salpeter}$, 
where the latter ratio is computed for the cutoff at 
$0.15 M_\odot$. Charlot, Worthey, \& Bressan (1996) have given a
compilation of mass-to-light ratios from various stellar
population synthesis calculations. They show that
$M/L_V$ is reasonably consistent among authors when 
the lower mass cutoff is fixed. From the Charlot et al.
(1996) compilation of population synthesis calculations, 
the GBF IMF, and the
conversion formula discussed above, we obtain
$(M/L_V)_{\rm GBF}= (4.0\pm0.3)+0.38(t_G-10{\rm Gyr})$, where
$t_G$ is the age of  the spheroid and the error represents the
model-dependent uncertainty. Using the Charlot {\it et al.}
(1996) synthetic calculation of the
$B-V$ color, we obtain  
$M/L_B=(5.4\pm0.3)+0.7(t_G-10{\rm Gyr})$. 
For $t_G =12\pm2$ Gyr we have  $M/L_V=4.0-5.9$, or
\beq M/L_B=5.4-8.3. \label{eq:3}\eeq 

The $M/L$ ratio in equation~(\ref{eq:3}) may be compared to 
the values inferred from the kinematics of the nuclear regions of
elliptical galaxies. van der Marel (1991) finds
$M/L_R=(6.64\pm0.28)h$ (Johnson $R$),  
which translates to $M/L_V=(7.72\pm0.33)h$ or 
\beq
	M/L_B=(9.93\pm0.42)h,\label{eq:3b}
\eeq
using $\langle V-R_J \rangle=0.68$,
corresponding to the average color $\langle B-V \rangle=0.92$ for
the 37 elliptical galaxies he used ($\langle B-V \rangle_\odot=0.65$ and
$\langle V-R_J \rangle_\odot=0.52$ are also used).  
There is excellent consistency between equations~(\ref{eq:3})
and~(\ref{eq:3b}) if the Hubble constant is in the range
$h=0.6-0.8$ and spheroid populations dominate the the mass
of these nuclear regions. 

Dynamical measures show $M/L$ depends on the luminosity in 
elliptical galaxies (Kormendy 1986): van der Marel (1991) 
finds $M/L=(L/L^*)^{0.35\pm0.05}$. This manifestation
of the color-magnitude relation of early-type galaxies 
probably reflects the effect of metallicity and perhaps also 
of dark matter halos. In the former case we estimate 
this dependence may reduce  the effective mean value of $M/L$
weighted by the contribution to the luminosity density by 17\%,
and accordingly reduce the lower end of the allowed range by
this amount.

We conclude that a good estimate of the mass-to-light ratio of
stars and their remnants in spheroids is
\beq 
	M/L_B({\rm spheroids})=6.5{+1.8 \atop -2.0}.\label{eq:3a}
\eeq
 
\subsubsection{Mass-to-Light Ratio of the Disk Component}

One way to estimate the mass-to-light ratio of the disk stars is
to use estimates  of the mass and luminosity column densities
 from surveys of stars and remnants in the solar neighborhood. 
Commonly used values are $\Sigma_M\simeq 45M_\odot$
pc$^{-2}$ (stars and stellar remnants) and $\Sigma_{L_V}\simeq 
15L_\odot$ pc$^{-2}$ (Binney \& Tremaine 1987; Bahcall \& Soneira 1980). 
GBF find that the mass column density must be reduced 
to $\Sigma_M\simeq 27.3M_\odot$~pc$^{-2}$ because of the turnover
of the luminosity function towards the faint end, as noted above.
With this new value we have $M/L_V
\simeq 1.82$ or $M/L_B\simeq 1.50$ for the disk with $B-V\simeq 0.44$
(de Vaucouleurs \& Pence 1978). This is not very different from
the early estimate by Mihalas \& Binney (1981), 
$M/L_B\simeq 1.2$.
 
The solar neighborhood value may be compared with $M/L_B$ from
population synthesis calculations. 
The Tinsley (1981) 10-Gyr model with a constant star
formation rate gives $M/L_B=1.2$ 
after a 20\% upward correction in the mass integral of the IMF
to match GBF.  
Shimasaku \& Fukugita's (1997) gas infall model (with effective
disk age  $\approx$ 10 Gyr)
gives $M/L_B=1.9$ after a 40\% reduction in $M$ to correct the IMF. 
 From a study of rotation curves for spiral galaxies,
Persic and Salucci (1992)
find $M/L_b=1.24h$. We conclude that there is reasonably small
scatter among these different approaches to the disk
mass-to-light ratio, and we adopt
\beq(M/L_B)_{\rm disk}=1.5\pm0.4. \label{eq:4}\eeq

There is good evidence that stars in gas-rich irregular (Im) 
galaxies are young. (For reviews see Fukugita, Hogan, \&\ Peebles
1996; Ellis 1997). If the mean age of irregular galaxy stars is
$\approx 5$~Gyr, 
the $B$ luminosity is 0.5 mag brighter and the mass in star
remnants is about 10\% smaller than for a 10 Gyr age disk. 
These corrections reduce $M/L$ in equation~(\ref{eq:4}) to
$M/L_B=1.1$ for irregular galaxies. A population synthesis model
for 8 Gyr age (Shimasaku \& Fukugita 1997) also gives $M/L_B=1.1$
after the correction to the IMF. We take 
\beq(M/L_B)_{\rm irr}=1.1\pm 0.25. \label{eq:5}\eeq

\subsubsection{Baryons in Stars and Remnants}

With the above numbers the mean mass densities in
our three classes of stars and their associated remnants are 
\beq
	\Omega_{\rm spheroid~stars}=\left( 0.00180{+0.00121 \atop 
	-0.00085}\right) h^{-1},\label{eq:6}
\eeq 
\beq
	\Omega_{\rm disk~stars}=\left( 0.00060{+0.00030 \atop
	-0.00024}\right) h^{-1}, \label{eq:7}\eeq
\beq
	\Omega_{\rm stars~in~Irr}=\left( 0.000048{+0.00033 \atop 
	-0.00026}\right) h^{-1},  \label{eq:8}
\eeq
in units of the critical density 
$\rho_{\rm crit}=2.775\times 10^{11}h^2M_\odot$~Mpc$^{-3}$. Here
and below the upper (lower) errors are obtained by taking all
errors in each step to maximize (minimize) the result. In this
sense the error ranges are conservative. 
The numbers in
equations~(\ref{eq:6}) to~(\ref{eq:8}) depend through the 
$M/L$ calculation on
age estimates, and hence depend on the Hubble
constant $h$ and other cosmological parameters in a complicated
way. If dynamics were used to estimate $M/L$ the estimates of
$\Omega$ would not depend on $h$, but in that case $h\approx 0.7$  
would be needed for dynamics to agree with the synthesis calculations.
Either way, consistency holds for $h\approx 0.7$ in a low
density universe.

\subsection{Atomic and Molecular Gas}

 From   HI 21-cm surveys  Rao and Briggs (1993) find 
$\Omega_{\rm HI}=(1.88\pm0.45)\times 10^{-4}h^{-1}$
(see also Fall \& Pei 1993, Zwaan et al. 1998).
%% This was eneterd as Zwann et all 1997; is 98 correct?
A 32\%\ addition (for 24\%\ helium mass fraction) to the
Rao-Briggs value  for helium yields  
%% I like Craig's remark that the label Omega_atomic is more logical
%% than omega_HI, though don't feel strongly. I've also changed the
%% label in eq. 35. But what do we do about clusters?
%% In a quick scan of Fabian and White I could find no
%% indication of a correction for helium (and am not inclined
%% to put one in at his late date!)
%% \beq\Omega_{\rm HI}=(0.00025\pm0.00006)h^{-1}.
%% \label{eq:9}\eeq
\beq\Omega_{\rm atomic}=(0.00025\pm0.00006)h^{-1}.
\label{eq:9}\eeq
This result is  quite secure because absorption line
studies show directly that most of the HI is in systems with
column densities high enough to detect in 21-cm emission.
These  columns are also generally high
enough to  resist photoionization so we are counting here
mostly regions dominated by neutral atomic gas. 

The ratios H$_2$/HI of mass in molecular and atomic hydrogen 
listed in Table~\ref{table1} for each morphological type 
are median values
 taken from the compilation of 
Young \& Scoville (1991), who base the molecular
hydrogen masses on CO observations. The representative masses in
atomic hydrogen in Table~\ref{table1} are from Roberts \&
Haynes~(1994). Combining these numbers with the distribution 
$\mu _t$ in morphology, weighted by HI mass,  we obtain 
the global ratio in galaxies
$\rho_{\rm H_2}/\rho_{\rm HI}=0.81$, and from equation~(\ref{eq:9}) 
\beq\Omega_{\rm H_2}=(0.00020 \pm 0.00006)h^{-1}.\label{eq:10}\eeq

\subsection{Baryons in Clusters of Galaxies}

The cluster mass function is approximated as (Bahcall \& Cen 1993)
\beq n_{cl}(>M)=4\times 10^{-5}h^3\Big({M \over M^*}\Big)^{-1}
\exp \Big(-{M \over M^*}\Big)~{\rm Mpc}^{-3}, \label{eq:11}\eeq
where $M^*=(1.8\pm0.3)\times10^{14}h^{-1}M_\odot$, and $M$ is the total
gravitational mass within a sphere of a radius of 
1.5$h^{-1}$Mpc (the Abell radius) centered on the cluster. 
%% Masataka, please check the revisions, which are meant only to
%% be stylistic. 
At the Abell radius the mass distribution is close to dynamical
equilibrium. The envelope of matter around the cluster at greater
distances merges into the ``large-scale structure'' and we take
this matter to be included in the field (as discussed in 
\S 2.4). 
We define clusters as  objects with mass $M>10^{14}hM_\odot$.
The integral $\int dM\, Mdn_{\rm cl}/dM$ gives the mean mass
density in clusters, 
\beq    \rho_{\rm cl}=\left( 7.7{+2.5 \atop -2.2}\right) \times 
	10^9h^2M_\odot\hbox{ Mpc}^{-3}.\label{eq:11ab}
\eeq
The contribution of this gravitational mass to the density parameter
is 
\beq\Omega _{\rm cl} = 0.028{+0.009\atop -0.008}.\label{eq:11a}\eeq

Intracluster plasma masses are well determined from X-ray
observations (Fabricant et al. 1986; Hughes 1989; 
White et al. 1993). The ratio of X-ray emitting gas mass to 
gravitational mass within the Abell radius from the survey by 
White \& Fabian (1995) is 
\beq
	(M_{\rm HII}/M_{\rm grav})_{\rm cl}=(0.056\pm 0.014)h^{-3/2}.
		\label{eq:11aa}
\eeq
In the White \& Fabian~(1995) sample this ratio shows no
correlation with cluster mass.
%% This fraction seems to depend little on the cluster mass.
The value (18) is independently verified by Myers et al. (1997),
 $(M_{\rm HII}/M_{\rm grav})_{\rm cl}=(0.061\pm 0.011)h^{-1}$
 from the measurement of the Sunyaev-Zeldovich effect in three clusters.
The product of equations~(\ref{eq:11a}) and~(\ref{eq:11aa}) is
\beq (\Omega_{\rm HII})_{\rm cl}=\left( 0.00155{+0.00100 \atop -0.00072} 
	\right) h^{-1.5}. \label{eq:12}\eeq
This is the contribution to the baryon budget by intracluster plasma.

%%??
Let us estimate the baryons in stars in galaxies in clusters.
Although these stars 
are included in the estimate for stars in \S~\ref{section2.1},
it is useful to compare their mass with the plasma mass, which we
can derive from 
the cluster mass-to-light ratio. In the discussion of the Coma
cluster by White et al. (1993),
straightforward applications of galaxy velocity dispersions or
the X-ray pressure gradient give 
$(M_{\rm grav}/L_B)_{\rm cl}\sim 370h$, or values as large as
$500h$ if the analysis is constrained by models from numerical
simulations of cluster formation. 
The CNOC value (Carlberg et al.
1996) transformed to the B band by $(B-r)_\odot=0.65$ and 
$\langle B-r \rangle= 1.03\pm0.1$ for the average color of S0
galaxies, after a passive evolutionary
correction for early type galaxies, is somwhat larger,
 $560h$. However, the CNOC luminosity density is correspondingly
 smaller than our
adopted value (equation (4)), and it is the product of this with
$M/L$ that matters in
deriving most global quantities; moreover
we suspect that the product  is 
more reliably estimated since in the end we are estimating
the luminosity density in cluster galaxy stars, which does
not depend on a mass estimate.
We therefore adopt a central value, 
\beq
	(M_{\rm grav}/L_B)_{\rm cl}=(450\pm 100)h.\label{eq:12a}
\eeq
For cluster galaxies (Dressler 1980)  we adopt the 
composition within the Abell radius
\beq
	\mu ({\rm E})=0.21,\qquad\mu({\rm S0})=0.44,\qquad
	\mu ({\rm S})=0.35.\label{eq:12aa}
\eeq
(This is somewhat richer in early-type galaxies than the
estimate given by Schechter \& Dressler (1987), but their sample
extends beyond the Abell radius.) The mass-to-light ratios
in equations~(\ref{eq:3a}) and~(\ref{eq:4}) and the bulge-to-disk
ratios in Table~\ref{table1} give $(M/L)_{\rm star}=4.5\pm1$. 
The ratio to equation~(\ref{eq:12a}) gives
$hM_{\rm stars}/M_{\rm grav}=0.010{+0.005 \atop -0.004}$.
The product with equation~(\ref{eq:11a}) is the contribution to
the density parameter by stars in clusters,
\beq
	(\Omega_{\rm stars})_{\rm cl}=
	\left(0.0003{+0.0003 \atop -0.0002}\right) h^{-1}, 
\label{eq:13}\eeq
about ten percent of the total in equations~(\ref{eq:6})
to~(\ref{eq:8}). 

\subsection{Baryons in Groups}\label{section2.4}

Plasma associated with galaxies outside the great
clusters makes a significant and still quite uncertain
contribution to the baryon budget. The
detected X-ray emission from plasma in groups is softer than
for clusters, and softest for groups dominated by late-type
galaxies. It is a reasonable presumption that the absence of
detections often represents a lower 
virial temperature, rather than the absence of plasma.

We follow two approaches in estimating the mass in plasma
associated with galaxies outside clusters. In the first, we start 
with a  characteristic value for the ratio of plasma mass to
gravitational mass in groups with X-ray detections. We apply this 
ratio to all galaxies, on the assumption that all field galaxies
are likely to have inherited a similar plasma mass,
to estimate the density in the warm phase  (\S~\ref{section2.4.1}). To this we
add an estimate from quasar absorption
 of the  plasma mass in low surface density cool
 clouds that certainly
are not counted in the X-ray measurements
(\S~\ref{section2.4.2}). The second 
approach (\S~\ref{section2.4.3}) assumes the
plasma-to-gravitational mass ratio is the same in clusters and
the field.

\subsubsection{ Warm Plasma in Groups}\label{section2.4.1}

Some surveys are available from low energy 
X-ray observations with ROSAT (Mulchaey et al. 1996). 
The X-ray emission from 18 groups with total mass ranging
 from $1.2$ to $8.3\times 10^{13}h M_\odot$ corresponds to plasma 
mass fractions ranging from $0.004h^{-3/2}$ (H97) to
$0.09h^{-3/2}$ (NGC 4261). The average is 
\beq
	(M_{\rm HII}/M_{\rm grav})_{\rm group}=(0.022\pm 0.005)h^{-3/2},
			\label{eq:14}
\eeq
significantly below the number for clusters
(eq.~[\ref{eq:11aa}]). In fact the baryon fraction shows a trend
increasing with the group mass, approaching to the cluster value
at the high mass end.
This could be because groups are
intrinsically poorer in plasma, or because much of the plasma is 
cooler and so escapes detection as an X-ray source. The latter is
in line with the shallower gravitational potential wells in
groups. The cool plasma clouds detected by Lyman-$\alpha$
resonance absorption (\S~\ref{section2.4.2}) similarly are not
detected as X-ray sources. Thus the plasma identifiable from its 
discrete X-ray emission might be considered a lower limit to the
net plasma associated with groups of galaxies. 

To convert equation~(\ref{eq:14}) into a mean baryon density we
need the mean gravitational mass associated with field galaxies
(which almost always are in groups).
The following measures may be compared. 
First, we can extrapolate the Bahcall-Cen (1993) mass function
(eq.~[\ref{eq:11}]), which they determined for 
$M>10^{13}hM_\odot$, to systems with the mass characteristic of
galaxies,  $M\sim 10^{12}hM_\odot$. The result of integrating this
mass function from $M=10^{12}hM_\odot$ to $M=10^{14}hM_\odot$ is
\beq\Omega=0.12\pm0.02.\label{eq:14.1}\eeq
The cutoff is the characteristic mass of an $L^\ast$ galaxy, the
minimum for a group. 

Second, we have dynamical mass measures from analyses of systems
of galaxies on scales smaller than about $10h^{-1}$~Mpc and
outside the rich clusters. The survey of results of these
analyses by Bahcall, Lubin \& Dorman (1995) indicates 
$M_{\rm grav}/L\simeq (200{+100\atop -50})h$. This with
the mean luminosity density in equation~(\ref{eq:1d}) gives 
\beq\Omega =0.14{+0.10\atop -0.05}.\label{eq:14.2}\eeq

Third, we can use the luminosity density,
scaling from   $M/L$ calibrated in the great clusters by taking
account of the difference in luminosities from the difference in
morphological mixes (see also Carlberg et al. 1997).
The small scatter in the 
color-magnitude relation for ellipticals and the spheroid
components of spirals suggests these stars formed early,  
so it is reasonable to assume that the ratio of spheroid luminosity
 to gravitational mass is 
 the same in clusters and the field. It would follow that the
cosmic mean mass-to-spheroid-light ratio is 
$(M/L_B)_{\rm  cos}=(M_{\rm grav}/L_B)_{\rm cl}
f({\rm sph,cos})/f({\rm sph,cl})= (270\pm 60)h$, where $f({\rm sph,cos})$
 is the cosmic mean spheroid population luminosity 
fraction (eq.~[\ref{eq:1}]), the value in clusters,
$f({\rm sph,cl})=0.64$, follows from
equation~(\ref{eq:12aa}), and $(M_{\rm grav}/L_B)_{\rm cl}$ 
is from equation~(\ref{eq:12a}). With 
equation~(\ref{eq:1d}) we have $\Omega =0.19{+0.07\atop -0.05}$. 
This however gives  the cosmic mean including
clusters; for  comparison to
equation~(\ref{eq:14.2}) we ought to convert to the value outside  
rich clusters. The ratio of equation~(\ref{eq:11ab}) to
equation~(\ref{eq:12a}) is the mean luminosity density
provided by the clusters, ${\cal L}_{\rm cl}=0.09{\cal L}$. 
The spheroid luminosity fraction in the field is
\beq
	f_{\rm field}({\rm sph})=({\cal L}f_{\rm cos}- 
	{\cal L}_{\rm cl}f_{\rm cl})/({\cal L} - {\cal L}_{\rm cl})
	=0.94f_{\rm cos}.\label{eq:14.2a}
\eeq
The assumption that the mass-to-spheroid-light ratio is universal  
thus indicates the density parameter in gravitational mass
outside the Abell radii of the great clusters is 
\beq
	\Omega =0.18{+0.07\atop -0.05}.\label{eq:14.3}
\eeq

Despite the substantial uncertainties in each of these arguments
 we are encouraged by the
consistency of equations~(\ref{eq:14.1}), (\ref{eq:14.2}), and
(\ref{eq:14.3}) to conclude that the density parameter in
gravitational mass that 
clusters with galaxies on scales $\lsim 10h^{-1}$~Mpc is likely
to be in the range
\beq
\Omega = 0.15{+0.10\atop -0.05},\label{eq:14.4}
\eeq
and that the spheroid-to-dark matter and baryon-to-dark matter
ratios indeed do not vary widely between clusters and the field.
As the evidence has indicated for some time (Peebles 1986)
and has been widely noted in recent years, this
low density parameter universe offers a natural interpretation of 
a variety of observations. Recent examples include the
age/distance scale relation, the abundance of cluster baryons
(White et al. 1993), the growth rate of correlation functions
(Peacock 1997),   the growth rate of the cluster mass function 
(Bahcall, Fan \& Cen 1997), and preliminary indications from
the distant supernova Hubble diagram (Garnavich et al. 1998, 
Perlmutter et al. 1998).

The product of equations~(\ref{eq:14}) and~(\ref{eq:14.4}) is 
the estimate of total plasma identified by X-ray
emission in groups,
\beq
	(\Omega_{\rm HII})_{\rm group}=\left( 
	0.003{+0.004\atop -0.002}\right) h^{-3/2}. \label{eq:14a}
\eeq
%% Masataka I changed -20 to 20 in the following line. And take note
%% of other alterations.
We remark that this estimate decreases only by $20$\% when
we incorporate explicitly the trend that the baryon fraction
increases as the group mass, ($M_{\rm HII}/M_{\rm grav})_{\rm group}
\sim 0.056h^{-3/2}(M_{\rm group}/0.6\times 10^{14}M_\odot)$ for 
$M_{\rm group}<0.6\times 10^{14}M_\odot$. As we have noted, an
alternative interpretation is that the trend is only apparent, a
result of less efficient detection of plasma around cooler
groups. 

\subsubsection{Plasma in Cool Low Surface Density
Clouds}\label{section2.4.2}

In addition to the hot plasma detected in X-rays,
there is a considerable amount of plasma in cool,
thermally stable, photoionized clouds with temperatures less than  
about 20,000~K. These cool low column density clouds are 
detected  by the 
Lyman-$\alpha$ resonance absorption line in quasar spectra.  They
 may be found as far
as 5 Mpc from a galaxy, and generally seem to avoid
both the voids and strong concentrations of galaxies 
 (Shull, Stocke, \&\ Penton 1996). This is 
consistent with the view that such clouds were left behind in
the gravitational assembly of groups of galaxies; those that
would have belonged to cluster members were incorporated and shock
heated by the assembly of the clusters, joining
the hot phase. In our budget for plasma
 in groups we thus use  the sum of the mean baryon densities in 
these clouds and in cluster baryons detected by X-ray
emission (eq.~[\ref{eq:14a}]).

The estimate of the baryon abundance in such clouds is 
very sensitive to model details
such as ionization and geometry.  Let us assume as an
illustrative  model that the
clouds are uniform density, isothermal spheres
 in photoionization
equilibrium, and the size is of the order of 100~kpc,
determined empirically by common absorption between
two nearby lines of sight.  
Then the mass of the clouds that yield neutral hydrogen column density 
$N_{\rm HI}$ are estimated, following Shull, Stocke \&\ Penton
(1996), to be 
\beq M_{\rm Ly\alpha}=7.1\times 10^8M_\odot
J^{1/2}_{-23}[T/(2\times 10^4)]^{0.375} 
  N^{1/2}_{14}(R/100 {\rm kpc})^{5/2} [(3+\alpha)/4.5]^{-1/2}
\label{eq:15}\eeq
where $J_{-23}$ is the ionizing UV flux at the Lyman limit in units of
$10^{-23}$erg s$^{-1}$cm$^{-2}$sr$^{-1}$Hz$^{-1}$, with a power index
$\alpha$, $T$ is the gas temperature, $N_{14}$ is the neutral hydrogen column
density in units of 10$^{14}$cm$^{-2}$, and $R$ is a typical
cloud size. The neutral hydrogen fraction $x=n_{\rm HI}/n_{\rm H}\simeq
6\times 10^{-5} J^{-1/2}_{-23}[T/(2\times 10^4)]^{-0.375}
  N^{1/2}_{14}(R/100 {\rm kpc})^{-1/2} [(3+\alpha)/4.5]^{1/2}$.
With the detection frequency of the absorbers at $z\simeq0$, 
$dN_c/dz=\phi_0
(\pi R^2)c^2/H_0\approx 86$, the mass density of the clouds is 
\beq
	\Omega_{Ly\alpha}={\phi_0M_c\over\rho_{\rm crit}}
	\simeq 0.00234 {J^{1/2}_{-23} [T/(2\times 10^4)]^{0.375}
	N^{1/2}_{14}(R/100 {\rm kpc})^{1/2}\over
	h [(3+\alpha)/4.5]^{1/2} }.\label{eq:16}
\eeq
This assumes spherical clouds.  If the clouds are oblate the density
decreases by the aspect ratio, which we assume to be between 1/5 and 1.

Direct simulations of
cloud formation and   absorption spectra permit a much more
detailed though model-dependent picture of the clouds. 
It has been demonstrated that these simulations give good
fits to the observations at $z\approx 3$
(see below) but they do not not yet yield useful approximations
to the situation at $z=0$. Thus the direct estimates of the
density parameter in this component remain quite uncertain.

\subsubsection{Scaling from Clusters to Groups}\label{section2.4.3}

As an alternative to adding the previous two directly observed 
components (warm X-ray emitting gas and cool Lyman-$\alpha$
absorbing gas), we can estimate the total plasma in groups
from the assumption that it has the same ratio to dark matter 
as in the rich clusters.
That is, the scaling used to arrive at the density parameter in
gravitational mass in the field 
 from spheroid luminosity (eq.~[\ref{eq:14.3}]) readily
generalizes to a scaling estimate of the density parameter in
plasma associated with groups.
 The product of
equation~(\ref{eq:11aa}) for the ratio of 
plasma to gravitational mass in clusters and
equation~(\ref{eq:14.4}) for the density parameter in the field
yields 
\beq
	(\Omega_{\rm HII})_{\rm field}=\left( 
	0.008{+0.010\atop -0.004}\right) h^{-3/2}. \label{eq:16a}
\eeq
In a calculation along similar lines Carlberg et al. (1997) find 
a similar result, 
$(\Omega_{\rm HII})_{\rm field}= (0.01 - 0.014) h^{-3/2}$.
The central value in equation~(\ref{eq:16a}) agrees with the upper
bound from the sum of the more direct estimates in
equations~(\ref{eq:14a}) and~(\ref{eq:16}). 

Equation~(\ref{eq:16a}) assumes a fair sample of plasma relative
to gravitational mass is assembled and preserved in the great
clusters.  If disks were less common in clusters because star
formation has been suppressed, equation~(\ref{eq:16a}) would
include baryons in disks in the field, but the correction is
small because the baryon mass in disk stars is small. It is not
thought  that cluster cooling flows or winds have seriously
depleted the cluster plasma mass. In a more direct argument
Renzini  (1997) finds that the iron and gas content
of clusters and groups indicate the clusters indeed have close to
the global star formation efficiency, and that the
corresponding enriched gas from groups has been ejected into
intergalactic space, presumably ending up in the form discussed
in \S~\S\ref{section2.4.1} and~\ref{section2.4.2}.

\subsection{Summary of the Low Redshift Budget}

Our estimates of the baryon budget at low redshift, expressed as
fractions of the critical Einstein-de~Sitter value, are listed in
lines~1 to~7 in Table 3. We rewrite the Hubble constant from
equation~(\ref{eq:1c}) as
$H_0=70h_{70}$~km~s$^{-1}$~Mpc$^{-1}$ to center it on a more
likely value. The central values in the table are computed from
the central values of the parameters entering the estimate. An
upper  bound is obtained by systematically choosing each
parameter at the end of its reasonable-looking range that
maximizes the contribution to the mean mass density, and likewise
for the lower bound.

Line 7a is an estimate of the X-ray-detected 
plasma around groups (eq.~[\ref{eq:14a}]), line 7b is an estimate
of the cooler component detected by resonance Lyman~$\alpha$
absorption in the trace HI fraction (eq.~[\ref{eq:16}]), and the
sum is an 
estimate of the plasma associated with groups of galaxies. Both 
entries are quite uncertain, and may well have missed a
substantial amount of cool plasma. Consistent with this, the
entry in line $7^\prime$ from the scaling from the plasma mass in 
clusters of galaxies (eq.~[\ref{eq:16a}]) is larger. We suspect
line $7^\prime$ is more reliable than the sum of lines~7a and~7b. 
In the sums over the 
budget we use line $7^\prime$ for the maximum and central values, 
and lines 7a and 7b instead of $7^\prime$ for the minimum.

The true values for each line are not likely to be close to the
maximum in all entries, or close to the minima in all entries,
but we retain the whole range in quoting the sums in each column
in line~8 because the sums are dominated by a few entries.

At $h_{\rm 70}=1$ our estimate of the baryons in stars
and remnants  is 
$\Omega =0.0035{+0.0025\atop -0.0014}$ times the
critical density. This star mass is only  about 17\% of our
sum over all detected baryons. In our budget, in the $2.2\mu$
$K$ band, 26\%\ of the light is from disk and irregular galaxies, 
in agreement with the fraction of the star mass. 
(We use $\langle B-K\rangle=4.27$ 
for ellipticals and bulges, and 2.78 for disks and irregulars).
In the $B$ band, 61\% of the mean luminosity density is from 
disks and irregular galaxies. This component represents only about 
25 \% of the star mass and only  about 4\% 
of our estimated sum of 
all observed baryons. 

\subsection{Uncounted Components}

We discuss here several mass components that are not
counted in our budget because the observational evidence
about them is still sketchy. 
	
\subsubsection{Warm Plasma in the Voids}

There could be a very large baryon mass in plasma in the voids.
Its emission may even have been observed: plasma 
at temperature $\approx 2\times 10^6$K  fits a component of the 
 diffuse soft X-ray
background (Wang \& McCray 1993). 
 This might be produced
locally by a relatively 
 small amount of gas in the Galaxy (for a summary of evidence for
and limits on
hot gas around the Milky Way see Moore \& Davis 1994), or it could be
produced by diffuse extragalactic gas with a mean density
\beq
(\Omega_{\rm HII})_{\rm IGM}=
	0.2\zeta^{-1/2}C^{-1/2}h_{70}^{-3/2},
		\label{eq:17}
\eeq
where the emissivity parameter ranges from $\zeta =1$  for solar
to $\zeta =0.1$ for primordial abundances, and the clumping parameter
$C\equiv \langle n_e^2\rangle /\langle n_e\rangle ^2$ may be
significantly greater than unity.
Gas at this temperature at low redshift has few other detectable
effects. For example,   the COBE limit on the Compton distortion
of the microwave background spectrum, $y<1.5\times 10^{-5}$
(Fixsen \etal 1996),
is produced by a Hubble length of gas at a mean
density $\Omega=(T/10^6)^{-1}h^{-1}$, so the bound on
the $y$-distortion is not useful here. On theoretical grounds however 
it is difficult to see how there could be much mass in  void
plasma. If primeval, it would be surprising that the voids contain
little else in the form of baryons, especially as detected
in diffuse helium absorption at $z\approx 3$ (see below).   If blown in,
it would be surprising if the plasma far exceeds the density of stars, and
the required temperature in this case (to achieve sufficient
gas velocity to refill the voids in a Hubble time) exceeds
$10^7$K, which does produce  excessive anisotropy and spectral
distortion in the background radiation.  

\subsubsection{MACHOs}

The massive compact halo objects (MACHOs) detected as
gravitational microlensing events could be stellar
remnants, that is, baryons. The nature of the MACHOs is not
known, and there is no secure estimate of their global abundance.
However, they do seem to comprise a new population not
otherwise accounted for.

Current results from the MACHO collaboration (Alcock et al. 1997)
indicate objects with mass comparable to that of the Sun 
(in one model, $0.5^{+0.3}_{-0.2}M_\odot$) may account for 20\%
to 100\%  of the dark mass in a standard spherical  halo between
the Milky Way and the Large Magellanic Cloud.
This result must be taken with caution, however, because the
experiment measures the MACHO mass column in just one direction
in one halo and the extrapolation to a global density is subject to
many uncalibrated assumptions.  The Galactic MACHO population 
may have asymmetries or the MACHOs may be concentrated relative
to the global dark matter in the halo. For example, the actual
mass of MACHOs inferred within 50kpc (based on a spherical model, but
without extrapolating to larger radii) is estimated  at 68\% confidence to be 
13 to 32$\times 10^{10}M_\odot$.
This can be  compared to 
the   mass of the disk (about $6\times 10^{10}M_\odot$).  
A reasonable estimate of the minimum global density of 
MACHOs, viewing them as a new Galactic stellar  population,
 is to assume that all disk galaxies have  the same
ratio of MACHO to disk mass, taking the low end of 
the estimated range; this yields 
\beq
 \Omega_{\rm MACHO, min}=  2.2 \Omega_{\rm disks,min}= 0.0011
h_{70}^{-1},
		\label{eq:17a}
\eeq
   a minor entry in the budget.
On the other hand the data are consistent with
all of the Galactic dark matter being in MACHOs,
so a reasonable upper limit derives from assuming that
MACHOs comprise 100\%\ of the density parameter in
gravitation mass in equation~(\ref{eq:14.4}), giving  
$\Omega _{\rm MACHO,max}= 0.25$, and making MACHOs the
dominant entry in the baryon budget. 
(At this
level one expects detectable effects due to microlensing
of quasar continuum emission regions, e.g. Dalcanton et al. 1994).

 Assessments of this serious
uncertainty will be guided by advances in observational
constraints on the nature and amount of the MACHOs, and perhaps
also by advances in understanding processes of star formation and
death that could produce a substantial mass in baryonic
MACHOs without leaving an unacceptably large amount of debris.

\subsubsection{Dwarf Galaxies and Low Surface Brightness Galaxies}

The fainter end of the galaxy luminosity function has been a matter
of debate for some time, but recent work with larger volume
surveys (Lin et al. 1996; Zucca et al. 1997) indicates the
luminosity function is close to flat for at least 5 mag down from $L^*$ 
for the local field galaxies. There is some evidence for
a rise of the fainter tail right after the shoulder of
$L^*$ (Ellis et al. 1996), but this is visible only
in a high $z$ sample selected in the blue band, 
for which small galaxies show extraordinary star formation
activity without contributing much to the baryon budget. 
This means the contribution of dwarf galaxies to the baryon
budget is small, and the estimate from the local luminosity 
density likely is more secure. A rise
in the luminosity function somewhat below $L^*$ 
for cluster members also has been reported. (For the most
recent literature see Phillipps et al. 1997). The contribution
to the mass density from these cluster dwarf galaxies cannot
be dominant unless this sharp rise continues to very
low mass, which is not likely since it would violate measurements
of cluster surface brightnesses. 

The review by Bothun, Impey, \&\ McGaugh (1997) indicates
the importance of galaxies with surface brightnesses
lower than the 
more readily detectable ``normal'' galaxies. It is not clear however that
low surface brightness (LSB) galaxies make a significant
contribution to the baryon budget. 
LSB galaxies contribute to the budget by their
stars and star remnants, neutral gas, and plasma. 
The 21-cm observations by Briggs~(1997) indicate there is not a
significant contribution to the mean 
density of atomic hydrogen from gas-rich LSB galaxies in the 
field at distances $\lsim 10h^{-1}$~Mpc. A direct constraint on
diffuse plasma around LSB galaxies is much more difficult, but
since LSB galaxies avoid the voids defined by the more
visible galaxies this is the problem of counting the plasma
concentrated around galaxies outside the great clusters, as
summarized in lines 7a, 7b, and $7^\prime$.  
As we have noted, diffuse baryons initially 
associated with LSB galaxies that end up as cluster members would
be counted in the X-ray measurements of 
$(\Omega _{\rm HII})_{\rm cl}$
(eq.~[\ref{eq:12}]) and in the scaled value for the plasma in the
field (eq.~[\ref{eq:16a}] and line 7$^\prime$ in Table~3).
The mean luminosity 
density in the LSB sample considered by Sprayberry et al. (1997)
is 15\%\ of our adopted value (eq.~[\ref{eq:1d}]).
The mean luminosity density from all stars not shrouded
by dust, including those in systems with luminosities or surface
brightnesses below detection thresholds, is constrained by 
measurements of the extragalactic contribution to the sky surface
brightness. Absolute measurements  in progress 
(Bernstein~1997)  are capable of reaching $\sim 3{\cal L}$
(eq.~[\ref{eq:1d}]), but  differential measurements
of background fluctuations
already constrain plausible new populations to 
contribute much less than $\cal L$ (Dalcanton et al. 1997,
Vogeley  1997). We cannot exclude the possibility that
there is a significant mass in brown dwarfs
or baryonic MACHOs in LSB galaxies, but we can note that 
since many of these galaxies seem to be in early states of
evolution they would not seem to have had much opportunity to
have sequestered mass in dark stars, and that 
the integrated baryon content in present-day LSB galaxies likely
is counted in the measures 
of diffuse gas at redshift $z\sim 3$, as discussed next. 

\section{The Baryon Budget at $z\approx 3$}\label{section3}

There has been significant progress in understanding stellar 
populations at high redshift (e.g., Madau et al. 1996), but many
issues remain open. It is fortunate for our purpose that stars
are subdominant in the budget at the present epoch and likely are
even less important at high redshift, so we can concentrate
on the constraints on diffuse gas from quasar absorption
line spectra. 

The neutral hydrogen at $z\sim 3$ is predominantly in the high
column density damped Lyman-$\alpha$ absorbers (DLAs; 
Lanzetta, Wolfe, \&\ Turnshek 1995 and references therein).
The amount of neutral hydrogen in the DLAs increases with
increasing redshift back to $z\sim 2$. Our adopted value for the 
the density parameter in this form at $2<z<3$ is from 
Storrie-Lombardi, Irwin, \&\ McMahon (1996):
\beq
   \Omega_{\rm atomic}(z=3)=(0.0013\pm 0.0003,0.0020\pm 0.0007)
(h/70)^{-1}.
		\label{eq:18}
\eeq
The two estimates are for $\Omega=0$
and $\Omega=1$ respectively, and include the mass of the accompanying
helium. Because these systems are optically thick
to ionizing radiation there is no correction for an ionized
fraction in the neutral gas.  (The significant
amount of mass
in plasma in HII regions around young stars or in regions exposed
to the intergalactic ionizing radiation
are included in the forest component below).  Equation~(\ref{eq:18})
could be an underestimate if extinction by dust in the gas
suppressed the  selection of quasars behind high column density
absorbers  (Fall and Pei 1993) or an overestimate if
gravitational lensing enhanced selection of lines of sight
through dust-free, gas-rich absorbers.

Yet another uncertainty is the residency time of gas in DLAs. 
The large velocities in the DLAs (Prochaska \&\ Wolfe 1997) 
indicate the cloud masses 
could be depleted by winds only if there were considerable energy
input from supernovae. The more likely scenario is that
the density parameter in DLAs is decreasing at $z\lsim 2$
because the HI is being converted into stars. The HI mass in DLAs
at $z =3$ is about half that in present-day stars (lines 1, 2, 3,
and 9 in Table~3). That could mean there is a significant mass in
stars in DLAs and other young galaxies at $z=3$ and/or that
intergalactic matter still is settling onto protogalaxies at
$z=3$. However, these issues do not affect the budget regarded
as a snapshot of conditions at $z=3$. 

The dominant baryonic mass component is the 
Lyman-$\alpha$ forest gas,  detected by the trace 
neutral hydrogen in plasma that fills space as a froth at $z=3$. 
The density can be estimated
 from CDM model simulations that give   good fits to
the observations  taking  into account  the distributions of cloud
shapes, sizes, flow velocities, and temperatures,
although  
these estimates are still
sensitive to the uncertain flux of ionizing radiation.  
Rauch et al. (1997) conclude that the baryon density parameter
needed to correctly reproduce the statistical
absorption (mostly due to   clouds at HI surface density
$\simeq10^{13\pm 1}$~cm$^{-2}$) is 
$\Omega _{\rm HII}h^2>0.017-0.021$,
depending on cosmological model. 
 Weinberg et al. (1997)  
quote a lower limit of $0.0125h^{-2}$.
Zhang et al. (1997), including a self-consistent analysis
of the ionizing spectrum, arrive at a range for 
$h=0.5$ of $0.03<\Omega_b<0.08$, with about half of this
in the forest clouds.
Smaller densities 
may be possible  however
  because the same absorption can be produced
by a  higher HI fraction, caused by higher
density contrast and lower
gas entropy (Wadsley and Bond 1996,
Bond and Wadsley 1997).
Because of the unresolved issues 
 we assign the current
estimates a ``B'' grade. We  
adopt in Table~\ref{table3} the range 
given by Zhang et al. (1997), scaling
by $h^{-3/2}$.  For the lower limit
we include not the total density but just that in the forest 
clouds; this is appropriate for consistency, since   the 
simulations predict a larger fraction of cooled baryonic matter
than we infer from observations.

Cool plasma between the forest clouds has lower density and
hence a lower neutral fraction and very low HI Lyman-$\alpha$
resonance optical depth. The amount of plasma in this form is best
probed by measurements of resonance absorption by the most abundant
absorbing ion,
singly ionized  helium (Jakobsen et al. 1994; Davidsen, Kriss, \&
Zheng 1996.) High resolution HST/GHRS quasar absorption line spectra
of two quasars now permit the separation of the more 
diffuse component of the HeII resonance absorption from 
the component in the HI Lyman-$\alpha$ forest clouds
(Hogan, Anderson, \&\ Rugers 1997,
  Reimers et al.  1997).  The upper bound on intercloud gas density
is derived
based on the maximum permitted  mean HeII Lyman-$\alpha$
optical depth allowed
after subtracting the minimal contribution from the
detected HI Lyman-$\alpha$ forest clouds, 
while adopting the hardest ionizing spectrum allowed
by the data, with opposite assumptions
leading to the lower bound. 
(Note that the upper bound is on photoionized cool gas;
 hot gas with thermally ionized helium is not constrained
by absorption, but is even less plausible at $z\approx 3$
than at $z\approx 0$.) 
  These limits are
 consistent with predictions from  numerical simulations of CDM
models, in which most of the gas
is clumped in redshift space by this time
 (Croft et al.  1997, Zhang et al. 1997, Bi and Davidsen 1997.)
Although in principle the  helium bounds are sound the present results
 are  given a ``B'' grade because they are as yet based on only
two quasars.

\section{Nucleosynthesis}\label{section4}

The observationally successful theory of the origin of the light
elements by nucleosynthesis at redshift $z\sim 10^9$ predicts 
the mean baryon density in terms of the primeval element
abundances (as reviewed extensively in the literature; for
example Walker \etal.  1991, 
Copi \etal 1995,  Hata et al. 1997).
We consider here standard homogeneous nucleosynthesis predictions for
abundances as a function of the baryon-to-photon ratio 
$\eta\equiv 10^{-10}\eta_{10}$, where the present baryon density
is $\Omega_{\rm baryon}h_{70}^2= 7.45\times 10^{-3}\eta_{10}$. In this
context the strongest constraints on $\Omega_{\rm baryon} $ derive from
the abundances of  helium and  deuterium.  

The primordial deuterium abundance is still uncertain
but we can quote reliable upper and lower bounds. A conservative
lower bound, $(D/H)_p \ge 2\times 10^{-5}$, comes from many sources---
the Jovian atmosphere (e.g. Niemann 1996), the interstellar medium
 (Ferlet \& Lemoine 1996,
 Linsky et al. 1995), and  quasar absorption lines (Tytler, Fan
\& Burles 1996), together
with the fact that no source outside the Big Bang  is known
to produce deuterium significantly. An upper bound,
$(D/H)_p \le 2\times 10^{-4}$, comes from several measurements
in  metal-poor quasar absorption systems
(Songaila et al. 1994; 
Carswell et al. 1994; 
Webb et al. 1997; Songalia, Wampler, \&\ Cowie, 1997).
Although the identification as deuterium in these 
systems is disputed
(Tytler, Burles, \& Kirkman 1997),  the lack
of  higher detected values,
 and the fact that significant D destruction
would normally be expected to produce significant metal enrichment,
makes this a robust upper limit.
 These yield the limits $1.7\le \eta_{10}\le 7.2$. The lower
bound is used for the minimum value of $\Omega_{\rm baryon}$ in
line~14 in Table~\ref{table3}.  For comparison we include the
central value favored by Burles and Tytler (1997),
 $(D/H)_p = 3.4\times 10^{-5}$, which yields
$\eta_{10}= 5.1$ or $\Omega_{\rm baryon}=0.039h_{70}^{-2}$.

The  helium abundance $Y$ is well measured in nearby
galaxies (e.g., Pagel et al. 1992; Skillmann et al. 1994;
Izotov, Thuan, \&\ Lipovetsky 1997). 
The primordial value $Y_p$ derived from these observations
depends on models of  
stellar enrichment, but the present datasets yield 
a nearly model-independent 2$\sigma$ Bayesian
upper limit $Y_p\le 0.243$ (Hogan, Olive, \&\ Scully 1997).
This corresponds to $\eta_{10}=3.6$, which we use for
the maximum value in line~13 in Table~\ref{table3}; it is about
half the upper limit from deuterium. Most current  studies
(e.g. Olive \& Steigman 1995) are consistent with the central
value $Y_p\simeq 0.23$ used in the table. 

\section{Discussion}\label{section5}

After a brief comparison with earlier
work (\S~\ref{section5.1}) we offer in \S~\ref{section5.2}
an assessment of the major 
uncertainties in our budget at low redshift, with emphasis on the
more indirect arguments we hope constrain the budget of baryons
in forms that are unobservable in practice. We review the test
from the theory for the origin of the light elements 
in \S~\ref{section5.3}, and in \S~\ref{section5.4} we compare
these results with measurements of the diffuse baryons at
$z\approx 3$. Some concluding remarks are offered in
\S~\ref{section5.5}. 

\subsection{Comparison with Earlier Estimates}\label{section5.1}

Let us compare our estimates with those by Persic
\& Salucci (1992), Gnedin \& Ostriker (1992), and  Bristow \&
Phillipps (1994). These three groups considered
baryons in stars (often classified into those in early and late
types of galaxies) and in hot X-ray emitting gas around clusters. 
The baryon abundance in stars estimated by Gnedin \& Ostriker 
is in good agreement with ours, and that by Persic
\& Salucci is close to our minimum estimate. On the
other hand, the estimate of  Bristow \& Phillipps is
6 times ours. The same pattern applies to baryons in X-ray emitting
gas: Persic \& Salucci obtained 1/3 our central value and   
Bristow \& Phillipps found twice our estimate. As a result
Persic \& Salucci find that the total amount of baryons is
substantially less than that expected from nucleosynthesis, while 
Bristow \& Phillipps find the nucleosynthesis
value is saturated by stars plus hot X-ray emitting
gas alone. Our attempt to take account of a 
broader range of constraints and states of 
baryons, including plasma in groups inferred
but not directly observed,   has led us to conclude that, apart
from baryons that may have been sequestered at very high
redshift, the bulk are either directly visible or in
forms which are well constrained and  accounted for
with plausible extrapolations.  

\subsection{Observable and Unobservable Baryons}\label{section5.2}

A baryon budget must be informed by an understanding of
what is reasonable and sensible within standard models for the   
physics, astronomy, and cosmology. We argue here that these
considerations coupled to the observations available now or
within reach significantly limit the possibilities for 
large gaps in the budget. 

{\it Chemically Bound Baryons}. 
Baryons confined by chemical bonds, as in comets, are
dark (apart from evaporation) at any significant distance, but we
know their contribution to the baryon mass within the luminous
parts of normal galaxies is not large: the mass in stars
accounts for the dynamics.
In the standard cosmology only a small fraction of primordial
hydrogen converts to molecular form. Some scenarios envisage
a significant mass in molecular clouds (Combes
and Pfenniger 1998), but it seems more likely to us
that the bulk of this material would by now have formed into stars,
as it has in the Galaxy.
  Baryons could be chemically bound to
heavier elements, but it is hard to see how 
this could be a significant mass component because it would
require prodigious heavy element production in closely confined
and closed systems and the radiated
nuclear binding energy would appear in the background
radiation. Besides these arguments we have the direct limits on the density
of the plentiful but inert gas helium which would separate from any
condensed component. In short, if the existence of a
significant mass in intergalactic comets could be established it
would contradict the standard cosmological model.
 The
far more credible proposition is that there is not a significant
mass fraction in chemically bound baryons.

{\it Dark Galaxies.}
The starlight from baryons gravitationally bound in known forms
of low surface brightness (LSB) galaxies is only about 15\% of
our adopted total. Thus if normal galaxies were an adequate guide
to their compositions the known LSB galaxies would be
insignificant reservoirs of baryons in stars. Discovery of reason
to suspect the contrary would be of great interest, of course. The
starlight in as yet undetected still lower surface brightness 
galaxies is constrained by the absolute surface brightness of the sky
and fluctuations thereof; a significant contribution
already appears unlikely from the smoothness of the background
in deep exposures. If measurements indicated the mean
optical extragalactic surface brightness
is larger than expected from known galaxies it certainly would
give reason to think we have missed an important baryon
component. 

If the surface brightness showed no such anomaly one
could still imagine there are dark galaxies with star populations 
dominated by brown dwarfs, but there are several arguments
against the idea.
There are high success rates in optical
identification of systems of normal galaxies with the objects
responsible for gravitational lensing events, for quasar
absorption line systems, and for X-ray sources. Even if massive
dark galaxies were assumed not to be capable of significant
lensing they could contain diffuse baryonic halos, so in this
scenario we would need to explain why they do not 
cause a significant rate of identification failures for quasar
absorption lines and X-ray sources.  
One way assumes the dark galaxies avoid the concentrations of
normal galaxies and the associated plasma, so they do not acquire
plasma halos, but there are two counter arguments. First,
in this picture there surely would be intermediate cases, dim
massive galaxies that avoid the normal ones, and there is no
evidence of them. Second, the familiar morphology-density
relation in observed galaxies goes the other way: gas-rich
galaxies prefer lower density environments, gas-poor anemic
spirals denser regions.

{\it Dust-Shrouded Stars.}
Populations of stars shrouded in dust are constrained by
the integrated far infrared background of reradiated starlight,
which is observed to have an integrated flux about twice
that from optical galaxies as estimated from the Hubble Deep Field
and is thus the repository of most of the nuclear energy released by stars
(Schlegel, Finkbeiner and Davis 1998; Hauser et al. 1998).
  The total mass in baryons in stars even  allowing for a
maximal  shrouded population must however still be small.
It is also interesting to compare the estimated total energy density 
with the corresponding global production of heavy elements.
Taking the total bolometric intensity of
starlight  to be about $50 {\rm\ nW\ m^{-2}  ster^{-1}}$ (based on Schlegel et al.'s fitted
value $32\pm13$ at 140$\mu$m), the energy density is
\beq
	u = 2\times 10^{-14} (1 + z){\rm erg\ cm^{-3}}= 1\times 10^{-8} (1 + z)
{\rm  MeV\ cm^{-3}}
\eeq
where the redshift factor corrects for energy lost after the
mean epoch of emission. Each nucleon that is burned to helium
releases 25 MeV in heat, 
rising to a total of about 30 MeV  per nucleon converted from hydrogen
to heavy elements. On dividing $u$ by 30 MeV we get the mean
nucleon number density in heavy elements produced in the
production of the  background light,
$ n_{\rm heavy} \sim 4\times 10^{-10} (1 + z) {\rm cm^{-3}}.$
Our central value for the baryon number density is
$n_{\rm baryons} = 1.1 \times 10^{-7} {\rm cm^{-3}}$; the
 ratio is the mass fraction in heavy elements to make the
 observed background, 
\beq
	Z \sim 0.004 (1 + z).
\eeq
If the bulk of the radiation were produced at z = 2 it would mean 
$Z$ is about 1 percent. Though not a precise constraint,
it is  a significant check that the number
roughly agrees  with the observed metallicity of baryonic
material.

{\it Black Holes}.
Baryons sequestered in black holes in the luminous parts of
normal galaxies are known to be subdominant to the baryons in
stars. It is quite unreasonable to imagine whole galaxies have 
been lost to relativistic collapse. Gas clouds at the Jeans
length at decoupling may be susceptible to relativistic collapse
but only a tiny fraction of material has small enough
angular momentum to form black holes, and 
even then  it is hard to imagine an appreciable
fraction of the cloud mass could collapse before the remainder is 
blown apart by radiation from the collapsing fraction.
Far before recombination even a tiny collapsing fraction    yields
 a significant mass fraction today,  although it 
requires either extremely large amplitude
perturbations in $\eta$, which tends to adversely affect the model for
light element production (e.g., Kurki-Suonio, Jedamzik, \&\
Mathews 1996),
or extreme fine tuning of the spectrum and amplitude of 
small-scale adiabatic perturbations  (Carr 1994). In these
scenarios baryons may be sequestered in compact objects at
redshifts $z\gsim 10^{10}$ that are in effect nonbaryonic for the
purpose of this paper. 
 
{\it Baryonic MACHOs}. Debris is a key issue for MACHOs interpreted
as star remnants. In known processes of formation of 
white dwarfs and neutron stars, winds and explosions disperse  
most of the original star  mass in
diffuse debris. Thus a $2M_\odot$ star leaves
a $0.5M_\odot$ remnant and $1.5M_\odot$ in debris, and the
ejected fraction is larger for larger stars. The lensing
observations are consistent with a density parameter 
$\Omega\sim 0.25$ in MACHOs, 
which would be a very considerable entry in the budget, but we
would have to explain what happened to the debris. 
The debris might be recycled and a large fraction ultimately
sequestered in many generations of MACHOs, but recycling 
seems unlikely in the low density halos of galaxies. Debris still
present as diffuse matter, with density parameter $\Omega\ge
0.014$, would violate  
constraints previously discussed on diffuse matter 
in groups and most models would also generate
excessive metals. The most plausible form of baryons is
brown dwarfs, though
this would require a second peak in the global IMF.
Brown dwarf masses are below the estimated mass range of
the observed MACHOs, but this could be the result of an
unfortunate distribution of velocities of the MACHOs in the 
direction of the LMC. Absent the discovery of observational
evidence for the brown dwarf picture, or the demonstration that
the formation of star remnants  
need not disperse much debris in observable forms,
we suspect the MACHO contribution to the
baryon budget is subdominant to the mass in plasma around groups. 

{\it Plasma Around Field Galaxies}.
There is a significant baryon density in stars in normal
galaxies, and a still larger density in plasma around the
galaxies. The latter is the largest and most uncertain entry in
our baryon budget at low redshift in Table~\ref{table3}. The
estimate in line~($7^\prime$) assumes the spheroid
star-to-gravitational mass  
ratio is the same in clusters and the field. If we have missed a
significant mass in spheroids, as in LSB galaxies, and the
missing fraction is the same in clusters and the field, it does
not affect the estimate. If spheroid star production were more
efficient in clusters it would mean that the estimated
group mass is too small and hence line~($7^\prime$) is too
small. If plasma were ejected (more than
spheroid stars) from clusters during assembly, line~($7^\prime$)
would be an overestimate. But neither of these
effects could be large without upsetting the condordance
between equations (24), (25) and (27). Comparison of the two estimates, 
line~($7^\prime$) and the sum of lines~(7a) and~(7b), shows a
factor of two difference, but it is easy to imagine that 
this is because the X-ray observations still miss considerable warm
plasma and we have not yet adequately modeled the cooler low surface
density clouds detected by absorption lines at low
redshift.

{\it Baryons in the Voids.}
There are galaxies in the low density regions, or voids, defined
by normal galaxies. Galaxies in low density regions tend to be
the later Hubble types, and so are more readily observable than
their counterparts in concentrations. Gas enrichment aside, no
known type of object prefers the voids. This includes dwarf and 
irregular galaxies, low surface brightness galaxies, and the
plasma clouds detected from absorption lines. The reasonable
presumption is that plasma clusters the way all other observed
baryons do, meaning there is not much mass in plasma in the
voids. The possible exception is gas with high enough pressure
to resist gravitational draining of the voids, but if the
density of such void plasma were comparable to that
estimated in groups it would have appeared already as excessive
helium absorption at $z\approx 3$. 

The lack of reasonable-looking alternatives leads us to  conclude that
a fair accounting of the the baryons is possible because most are
in states that can be observed or reliably constrained by
more indirect arguments. We now argue that there is a reasonable
case for the net baryon density parameter 
\beq 
	\Omega _{\rm baryon} = 0.021\pm 0.007, \label{eq:last}
\eeq
close to the central value of the sum in line~8 in
Table~\ref{table3}.

\subsection{The Nucleosynthesis Check}\label{section5.3}

The central density in equation~(\ref{eq:last}) with $h_{70}=1$
corresponds to $\eta_{10}= 2.8$. At this value of the
baryon-to-photon ratio the standard model for light element 
nucleosynthesis predicts  $Y_p(^4{\rm He})\simeq 0.238$ 
and D/H$\simeq 9\times 10^{-5}$.  This is 
consistent with  observations of the helium
abundance, if somewhat above the central value,
 and  with  current observations of the deuterium abundance, 
if somewhat toward the high end of the 
range of estimates  
(Songaila 1997).   The mean of 
 the upper limit from helium and the lower limit
from deuterium is  $\eta_{10}= 2.7$ which we adopt  as the central
value in line~14  of Table~3.

If the baryon density were at the low end
of our range of estimates in line~8 in Table~3, 
$\Omega _{\rm baryon} = 0.007$ ($\eta _{10}=1.0$ at $h_{70}=1$),
the standard model prediction for the deuterium abundance would be
unacceptably large. This could be remedied by assuming a
significant baryon mass has been sequestered in the 
MACHOs, but for the reasons explained above we are inclined to
suspect rather that the adopted minima in 
lines~7a and~7b in Table~3 seriously underestimate the mass in
 plasma. The baryon density at the high end of our
estimates is $\Omega _{\rm baryon} = 0.041$ ($\eta _{10}= 5.5$ 
at $h_{70}=1$). This density is consistent with the bound from the
deuterium abundance but it predicts $Y_p= 0.247$, well outside
the accepted bound on the primeval helium mass fraction. 
In our opinion the
likely interpretation is that our upper limit in line~8 of
Table~3 also is overly conservative. 

Our interpretation of the baryon budget assumes the standard
homogeneous model for primordial nucleosynthesis. An
inhomogeneous primeval entropy per baryon, with the appropriate
coherence length (Jedamzik \&\ Fuller 1995),
would require serious attention if further observations
confirmed that there are significant variations in the deuterium
abundances in high redshift HI clouds or that the helium
abundances in some dwarf galaxies are lower than that predicted
by the global baryon budget applied to the homogeneous
nucleosynthesis model. 

\subsection{The Check from the Budget at $z\approx 3$}\label{section5.4}

Stars are subdominant in the budget at low redshift 
and likely are even less significant at redshift 
$z\approx 3$, so it makes sense to compare the sum  of 
the high redshift diffuse gas components in Table 3
to the sum at low redshift and the prediction of light
element nucleosynthesis. 

The sum of the central values for diffuse baryons at 
$z\approx 3$ is $\Omega = 0.04$. This is consistent with the
maximum sum in our low redshift budget but again contradicts the
standard model for helium abundance. The sum of the 
minimum values at $z\simeq 3$ is within the error flags in
equation~(\ref{eq:last}) and the standard model for the light
elements. Thus there is no contradiction with
equation~(\ref{eq:last}), but advances in the observations of the
intergalactic medium at high redshift may provide a critical
test. 

\subsection{Concluding Remarks}\label{section5.5}

We have emphasized three themes. First, a robust
estimate of the present baryon budget is of central importance in
serving to clarify and sharpen issues of research in present-day
cosmology. Second, there is no guarantee that physical processes
in the formation and evolution of structure will have conspired to
place a significant fraction of the baryons in forms we can hope
to detect by their emission or absorption of radiation or
otherwise infer in a convincing way. Third, the developing
evidence is that Nature in fact has been kind: we 
may actually be able to arrive at a close to complete budget 
of the baryons in the observable universe.  

The value for $\Omega _{\rm baryon}$ in equation~(\ref{eq:last}) 
suggests or is consistent with three lines of 
ideas. First, it suggests baryonic MACHOs 
make an insignificant contribution to the global budget. 
Perhaps the MACHOs are  non-baryonic entities formed in the early
universe, or perhaps like the stellar components of the Milky Way
galaxy they 
are more concentrated to the inner halo than is nonbaryonic dark
matter. The second idea is that there are relatively few 
baryons in the voids defined by the galaxies, likely
because the baryons were swept out of the voids by
gravity. Gravity would gather low pressure dark matter with
the baryons and galaxies, implying that the mass density in
matter than can cluster is well below the Einstein-de~Sitter
value. This is consistent with our estimate of the total density
parameter in equation~(\ref{eq:14.1}). The third idea is that the
bulk of the forest baryons at $z=3$ have ended up now in warm
plasma around the galaxies outside the great clusters.
The mass in neutral hydrogen in the DLAs is about half that 
needed to account for the present-day mass in stars. We know
some mass already is in stars at $z=3$; formation of the
rest of the star mass at $z<3$ would be a modest drain on the mass
in the forest. 

An interesting feature of the budget is that
all stars and their remnants, together with cold phases
of collapsed gas, comprise a small fraction even of the small
fraction of the matter which is baryonic (Gnedin \& Ostriker
1992; Persic \& Salucci 1992). 
Just the 0.3\% of the baryons in irregular
galaxies shine brightly enough at $z=0.4$ to~1
to make an important
contribution to the rapid increase of $B$ band number counts of 
galaxies with increasing apparent magnitude.
 It would not be difficult in these circumstances to imagine 
that the distribution of the starlight has little to do with the
mass. But the consistency of the estimates of 
dynamical estimates of the total density parameter, as in
equations~(\ref{eq:14.1}), (\ref{eq:14.2}), (\ref{eq:14.3})
 suggests starlight 
nevertheless is a useful tracer of mass on the scale of the
distance between galaxies. 

The baryon budget still is quite uncertain: the sums in line~8 
in Table~\ref{table3} for the maximum and minimum low redshift
budget differ by a factor of six. We suspect this is largely a
result of an underestimate in line~(7a) for plasma in groups; 
improved information on soft X-ray emission would be of
considerable help. Perhaps improvements in the observational
constraints will confirm the tentative concordance in
equation~(\ref{eq:last}); perhaps better observations will reveal
an inconsistency that shows we have to reconsider some aspect of
the standard concepts of astronomy and cosmology.

\acknowledgements

We are grateful to J. Bahcall for detailed discussions of the
local star luminosity function and to R. Carlberg,
J. Stadel, C. Stubbs, and E. Vishniac  for useful
comments. This work was supported in part by
a Grant-in-Aid of the Ministry of
Education of Japan (09640336) and the Japan Society for the 
Promotion of Science, by NASA and the NSF at the
University of Washington, and by the NSF at
Princeton University. M.F. acknowledges support from
the Fuji Xerox Corporation at Princeton.

\clearpage

\begin{deluxetable}{lcccccc}
\tablecolumns{7}
\tablewidth{0pc}
\tablecaption{Galaxy Parameters\label{table1}}
\tablehead{\colhead{} &\colhead{E} & \colhead{S0} & \colhead{Sab} & 
\colhead{Sbc} & \colhead{Scd} & \colhead{Irr}}
\startdata
bulge fraction $\kappa({\rm r})$
&   1     &  0.75  &   0.40   &   0.24  &   0.10  & 0  \cr
bulge fraction $\kappa({\rm B})$ 
&   1     &  0.64  &   0.33   &   0.16  &   0.061 & 0  \cr
morphology fraction $\mu ({\rm B})$ &  0.11  &  0.21  &  0.28   &   0.29  &  
0.045 & 0.061 \cr 
HI typical mass (10$^9M_\odot)$ 
 &  0  &  0.64  &   2.5   &  4.0  &  2.9 &  1.8 \cr
H$_2$/HI &  \nodata &  2.5   &  1.0  &  0.63  & 0.25    & 0.06 \cr
\enddata
\end{deluxetable}

\bigskip

\begin{deluxetable}{lcccc}
\tablecolumns{5}
\tablewidth{0pc}
\tablecaption{Luminosity Function Parameters\label{table2}}
\tablehead{\colhead{Author} &\colhead{$M^*({\rm B})$} & \colhead{$\alpha$} 
& \colhead{$(\phi ^\ast$)\tablenotemark{a} 
} & \colhead {$\cal L({\rm B})$}\tablenotemark{b} }
\startdata
Efstathiou et al. (1988) & $-19.68$   &  $-1.1$   & 1.56 &  1.93 \cr
Loveday et al. (1992) &  $-19.50$  &   $-0.97$   &  1.40  &  1.35 \cr
da Costa et al. (1994)   &  $-19.5$  &   $-1.20$   &  1.5  &  1.71 \cr
Marzke et al. (1994) & $-18.8$ & $-1.0$ & 4.0  &  2.1  \cr
Lilly et al. (1995)   & $-19.36$ &  $-1.03$  & 2.48   & 2.18  \cr
Ellis et al. (1996)   &  $-19.20$  & $-1.09$  & 2.6  &  2.05 \cr
Zucca et al. (1997)  &  $-19.61$  &   $-1.22$  &  2.0  &   2.2 \cr
\enddata
\tablenotetext{a}{Unit $=0.01h^3\hbox{ Mpc}^{-3}$ }
\tablenotetext{b}{Unit $=10^8hL_\odot\hbox{ Mpc}^{-3}$ }
\end{deluxetable}

\clearpage

\begin{deluxetable}{lllllc}
\tablecolumns{6}
\tablewidth{0pc}
\tablecaption{The Baryon Budget\label{table3}}
\tablehead{\colhead{} & \colhead{Component} &\colhead{Central} & 
\colhead{Maximum} & \colhead{Minimum} 
& \colhead{Grade}\hskip -2truemm\tablenotemark{a} }
\startdata
  & Observed at $z\approx 0$:\cr
1 & stars in spheroids &$ 0.0026 h_{70}^{-1}$&  $0.0043 h_{70}^{-1}$
      &   $0.0014 h_{70}^{-1}$  &A  \cr
2 & stars in disks & $0.00086 h_{70}^{-1}$&  $0.00129 h_{70}^{-1}$ 
&  $0.00051 h_{70}^{-1}$  &A--  \cr
3 & stars in irregulars & $0.000069 h_{70}^{-1}$ & $0.000116 h_{70}^{-1}$
& $0.000033 h_{70}^{-1}$  &B  \cr
4 & neutral atomic gas & $0.00033 h_{70}^{-1}$& $0.00041 h_{70}^{-1}$ 
&  $0.00025 h_{70}^{-1}$ &A  \cr
5 & molecular gas & $0.00030 h_{70}^{-1}$& $0.00037 h_{70}^{-1}$ 
& $0.00023 h_{70}^{-1}$ &A-- \cr
6 & plasma in clusters & $0.0026 h_{70}^{-1.5}$& $0.0044 h_{70}^{-1.5}$
  & $0.0014 h_{70}^{-1.5}$&A   \cr
7a & warm plasma in groups & $0.0056 h_{70}^{-1.5}$ 
&  $0.0115 h_{70}^{-1.5}$  & $0.0029 h_{70}^{-1.5}$  &B  \cr
7b & cool plasma & $0.002 h_{70}^{-1}$& $0.003 h_{70}^{-1}$ 
& $0.0007 h_{70}^{-1}$   &C \cr
$7^\prime$ & plasma in groups & $0.014 h_{70}^{-1}$ & $0.030 h_{70}^{-1}$
&  $0.0072 h_{70}^{-1}$ &B \cr
\cr \cline{1-6}\cr
8 & sum (at $h=70$ and $z\simeq 0$)
	&  0.021   &  0.041   &  0.007   &  \nodata   \cr
\cr\cline{1-6}\cr 
& Gas components at $z\approx 3$: \cr
9 & damped absorbers & $0.0015 h_{70}^{-1}$   &  
$ 0.0027 h_{70}^{-1}$
      &   $0.0007 h_{70}^{-1}$  &   A--  \cr
10 & Lyman-$\alpha$ forest clouds & $0.04 h_{70}^{-1.5}$  & 
$0.05 h_{70}^{-1.5}$ &  $0.01 h_{70}^{-1.5}$  &   B  \cr
11 & intercloud  gas (HeII) & \nodata &  $0.01 h_{70}^{-1.5}$
& $0.0001 h_{70}^{-1}$  &B  \cr
\cr\cline{1-6}\cr
 & Abundances of:&   &   &   &  \cr
12 & deuterium & $0.04 h_{70}^{-2}$  & $0.054 h_{70}^{-2}$ & $0.013 h_{70}^{-2}$ & A \cr
13 & helium& $0.010 h_{70}^{-2}$ & $0.027 h_{70}^{-2}$ &   & A  \cr
\cr\cline{1-6}\cr 
14 & Nucleosynthesis 
& $0.020 h_{70}^{-2}$ & $0.027 h_{70}^{-2}$ & $0.013
h_{70}^{-2}$ & \nodata  \cr 
\enddata
\tablenotetext{a}{Confidence of evaluation, from
A (robust) to C (highly uncertain)}
\end{deluxetable}

\end{document}